# Feed drive modelling for the simulation of tool path tracking in multi-axis High Speed Machining

## David Prévost and Sylvain Lavernhe

Laboratoire Universitaire de Recherche en Production Automatisée ENS Cachan
61 avenue du Président Wilson
94235 Cachan cedex, France
E-mail : david.prevost@lurpa.ens-cachan.fr
E-mail : sylvain.lavernhe@lurpa.ens-cachan.fr

## Claire Lartigue

Laboratoire Universitaire de recherche en Production Automatisée, ENS Cachan
61 avenue du Président Wilson
94235 Cachan cedex, France
IUT Cachan, Université Paris Sud 11
9 Avenue de la division Leclerc
94234 Cachan cedex, France
E-mail : claire.lartigue@lurpa.ens-cachan.fr

## Didier Dumur

Control Department, SUPELEC
3 rue Joliot Curie
91192 Gif-sur-Yvette cedex, France
E-mail: didier.dumur@supelec.fr

**Abstract:** Within the context of High Speed Machining, it is essential to manage the trajectory generation to achieve both high surface quality and high productivity. As feed drives are one part of the set Machine tool - Numerical Controller, it is necessary to improve their performances to optimize feed drive dynamics during trajectory follow up. Hence, this paper deals with the modelling of the feed drive in the case of multi axis machining. This model can be used for the simulation of axis dynamics and tool-path tracking to tune parameters and optimize new frameworks of command strategies. A procedure of identification based on modern NC capabilities is presented and applied to industrial HSM centres. Efficiency of this modelling is assessed by experimental verifications on various representative trajectories. After implementing a Generalized Predictive Control, reliable simulations are performed thanks to the model. These simulations can then be used to tune parameters of this new framework according to the tool-path geometry.

## 1 Introduction

High-Speed Machining (HSM) is now a standard process for the machining of sculptured surfaces as it allows high productivity combined with a good geometrical and dimensional accuracy (Krajnik and Kopač, 2004). However, the numerous stages of the process, from the tool trajectory computation to its execution on the machine tool, involves a large number of parameters, data transformations and information exchanges, which may affect global productivity and final part quality (Schmitz et al., 2008). Generally, the process is divided in three main stages (Figure 1): the CAM stage, the NC trajectory generation stage, and the cutting process itself (Tournier, 2005).

**Figure 1** Process stages

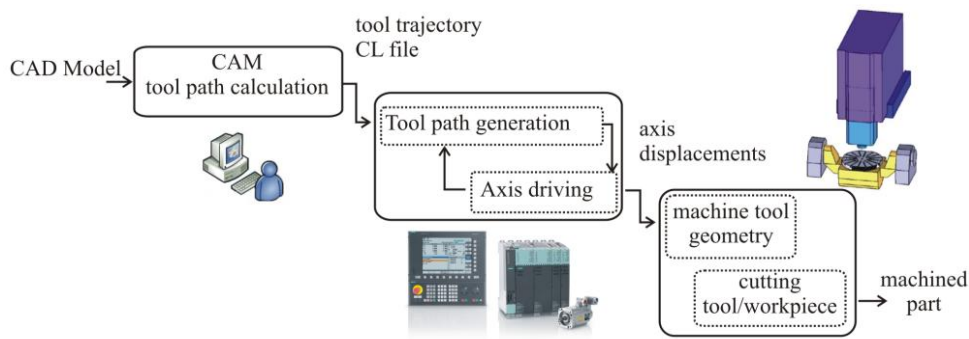

The CAM stage transforms the CAD model into a trajectory according to the machining strategy giving the CL-file. Sources of errors are mainly due to the calculation methods of tool positioning, and to the approximation of the CAD model by a set of points. The use of polynomial trajectory limits such errors (Lartigue, 2004).

Starting from the tool trajectory defined in the CL-file, the second stage carries out axis displacements. This stage calculates of the axis command followed by the trajectory tracking by the feed drives. During the last stage, the actual tool movement relatively to the part surface is executed on the machine tool; the cutting process thus generates the final machined surfaces. Therefore, the second stage is a key stage as it ensures the link between the numerical domain of the tool trajectory calculation and the physical domain of tool path execution by the axes. As such, it deserves a detailed description.

Dugas proposes a description of the process through three main levels defined in Figure 2 (Dugas, 2002). The level 1 (or numerical level) interprets in real time the program lines of the CL-file to generate axis commands. In this level, the NC unit calculates the position and velocity set points based on the kinematics limits of each axis. The level 2 (or *Feed drive level*) corresponds to axis cards and drives which perform axis commands, after the conversion of analogue data into digital data. The level 3 (or *kinematics level*) consists of the mechanical structure of the machine tool. The mechanical chain creates the relative movement between the tool and the part from the axis commands previously calculated. This last level is the entry point of the cutting stage.

**Figure 2** Global architecture of a NC machine tool

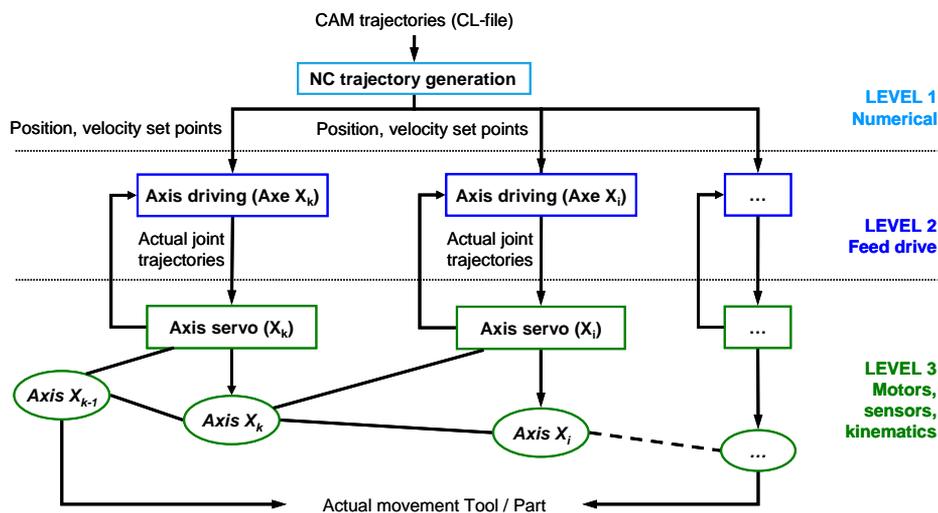

Within the context of virtual machine tool, the prediction of the global process behaviour relies on the modelling of each level (Altintas et al., 2005). Most authors focus on a specific part of the modelling. In a previous work, a first model of trajectory execution has been proposed (Lavernhe et al., 2008a-2008b). By integrating kinematics performance, this model allows the simulation of the actual tool path generation during machining by determination of position, velocity, acceleration and jerk of each axis. Results have in particular highlighted that the step of interpolation (level 1) is source of high decreases of the velocity along the tool path, and geometrical approximations involving deviations on the machined part. However, in order to improve the simulation with respect to the real behaviour, a modelling of the feed drives (level 2) separated from the modelling of the interpolator is necessary. It would thus be possible to predict from the set points calculated at the first level the actual tracking of the trajectory resulting from the feed drives and, consequently, the analysis of the tracking and contour errors (Ramesh et al., 2005).

The paper deals with the modelling of the feed drives (level 2) with the aim of predicting actual axis displacements. With such a model, it becomes possible to predict the final geometrical error as the sum of the errors related to each level.

Many studies for modelling of feed drives have been carried out in the literature (Whalley et al., 2005, Yeung et al. 2006b, Barre et al., 2002). Most of these works are performed on open CNC architectures or on specific machine tools (robots or agile machines), and in most cases considering one or two translational axes only. In the paper, the axis drive model proposed is deliberately simple (but realistic) to be applied to any type of axis (translation or rotation) and whatever the industrial high-speed machine tool. The model is developed for the context of HSM, considering finishing conditions so that disturbances due to cutting loads can be neglected. The model is assessed through various examples involving 2D or 3D trajectories. The result is an efficient model of the complete servo system, including existing axis controllers on industrial machining centres. Moreover, as the model proposed is generic, the implementation of various command strategies such as Generalized Predictive Control (GPC) is possible. To illustrate this aspect, the design and tuning of a Generalized Predictive Control (GPC) structure is proposed and implemented for the position loop of each axis drive.

## 2 Dynamic model

Common feedback structures used in industrial CNC drives are often based on a cascaded control structure. A typical cascaded structure (Figure 3) consists of several nested loops with dynamics becoming faster from the external to the inner loops.

**Figure 3** Axis control structure

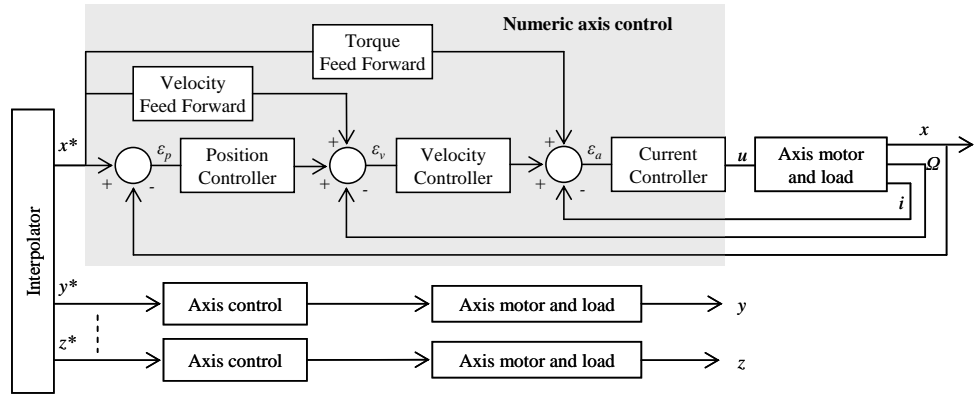

The current inner loop aims at controlling the motor torque as the torque is directly proportional to the armature current. Within the cascaded framework, its high bandwidth enables fast disturbances rejection. The current loop is closed using a proportional integral (PI) control. The velocity central loop is also closed using a PI controller. The position loop is the most external one with a larger sampling period. This last loop is closed using a proportional gain. Velocity and torque feedforward actions are added to the structure with the aim of reducing the tracking error resulting from the lack of integrative action in the position controller. The structure defined above is commonly used in industrial CNC drives with the major drawback that the structure is not open, essentially for historical and robustness reasons.

The proposed model relies on the feed drive structure presented in Figure 3. Axes are modelled in an independent way, neglecting the existing coupling interactions. Some studies focus on these phenomena but they concern specific machines such as welding robots (Ramesh et al., 2005) or agile structure machines (Barre et al., 2002). Besides, the model assumes the joints to be perfect and the components are supposed to be non deformable rigid bodies. The complete and generic model is built for translational axes as well as rotational ones. The implementation of the model is performed in the Matlab/Simulink™ environment (see Appendix 1). Depending on the machine architecture, some axes can be affected by gravity. This effect is taken into account by adding a resistant load (force or torque) to the motor. This resistant load is constant for a vertical axis whereas it depends on the axis angular position for a tilting table. Moreover, system of gravity compensation may exist. It must be modelled depending on technological solutions chosen by the manufacturer. Considering a context of finishing operation, disturbances due to cutting forces are neglected compared to dynamical effects (inertias, masses).

As the model must be generic and simple, some physical linked to the mechanical behaviour such as stiffness of the joints, backlashes, vibrations, are not modelled at this step (Siemens®, 2002; Whalley et al., 2005; Yeung et al., 2006a). Numerical filters and numerical compensations are aggregated into three adjustment delay parameters, ($\alpha$, $\beta$ and $\gamma$) set in both the position and the feedforward loops. For each axis, the input of the model is the setpoint calculated by the interpolator and transmitted to the axis cards, so-called Position Setpoint at Entrance of the Controller (PSEC). The choice of this setpoint is adopted in order to be free from numerical operations that are specific to each manufacturer (Bloch et al., 2001; Siemens®, 2002). The main output of the model is the simulated position (SP). Other inner variables of the model, such as the Simulated Velocity (SV) or the Simulated Motor Current (SMC) can be collected during simulation. In the next section, the identification step of the model parameters is detailed.

This model could be declined for different types of machine tool, with different motorization technologies. In the paper, the identification method is applied to one machine tool, which is a 5 axis Mikron UCP 710 HSM centre located in the LURPA lab, equipped with a RRTTT structure, rotary synchronous motors and Siemens Sinumerik 840D power line. Only the application to this 5-axis milling centre with a classical motorization is developed in the following sections.

## 3 Model parameter identification

### 3.1 Identification process

Generally, all the feedback controller parameters, such as the PI controllers, the proportional gain or the sampling periods are given by the CNC documentation. The parameters to be identified are thus the equivalent inertia, the friction coefficients and the feedforward compensations. In literature, many techniques for parameter identification exist. Some of them rely on frequency analysis by disconnecting the servo loops (Erkorkmaz and Altintas, 2001; Yeung et al., 2006a). Global techniques are also proposed so that the overall closed-loop dynamics is determined by running a standardized G-code and capturing the commanded and measured positions (Erkorkmaz and Wong, 2007). However, these techniques can only be easily performed for open architecture machines, which is not the case for industrial machining centres. Modern CNC systems offer the possibility to capture various variables in real-time during the trajectory execution, such as current, position, velocity and acceleration. This functionality is used during the identification step but also for the determination of the feedforward actions (e.g. Torque Feed Forward Set point TFFWS and Velocity Feed Forward Set point, VFFWS).

## 3.2 Motor modelling

Motors, either linear or rotational, can be modelled by direct current motors, which are sufficiently precise as suggested in (Boldea and Nasar, 1997). Typical equations for direct current motors are given in (1). They are transformed in the Laplace domain to be implemented in the proposed model (see Appendix 1).

$$\begin{cases} E(t) = K_e . \Omega_m(t) \\ U(t) = R.i(t) + L. \dfrac{d(i(t))}{dt} + E(t) \\ C_m(t) = K_t . i(t) \\ J_{eq} . \dfrac{d(\Omega_m(t))}{dt} = C_m(t) - C_r(t) \end{cases} \quad (1)$$

$E$ is the counter electromotive force, $\Omega_m$ the instantaneous rotation velocity, $K_e$ the constant of counter electromotive force. $U$ is the voltage across armature terminals, $R$ is the electric resistance, $L$ is the inductance, $i$ the current. $C_m$ stands for the motor torque, and $K_t$ is the torque constant. $J_{eq}$ is the equivalent inertia projected on the motor axis and $C_r$ gathers all resistant torques.

## 3.3 Friction modelling

In axis control, in the absence of cutting forces, the dominant source of disturbance is friction. Several studies lead to the modelling of friction through relationships between friction, position, speed and temperature of rigid bodies (Kikuuwe et al., 2005; Olsson et al., 1998). In this paper, a friction model depending only on axis velocity is chosen, as suggested by (Olsson et al., 1998) and (Mennon et al., 1999). The friction model is the combination of both viscous and Coulomb friction. When acceleration is zero, the motor torque is equal to the resistant torque. Considering equations (1), this yields to:

$$C_m(t) = C_r(t) = K_t . i(t) \quad (2)$$

If the resistant torque is only caused by friction, the equivalent current called $i_{friction}$ can be identified through NC recordings. Indeed, it is possible to evaluate the actual friction law by recording the current $i_{friction}$ for various displacements with constant velocities. Trials performed on the machine tool show that $i_{friction}$ is independent of the axis position and only depends on the velocity (Figure 4).

**Figure 4** Current and speed measurement (X axis)

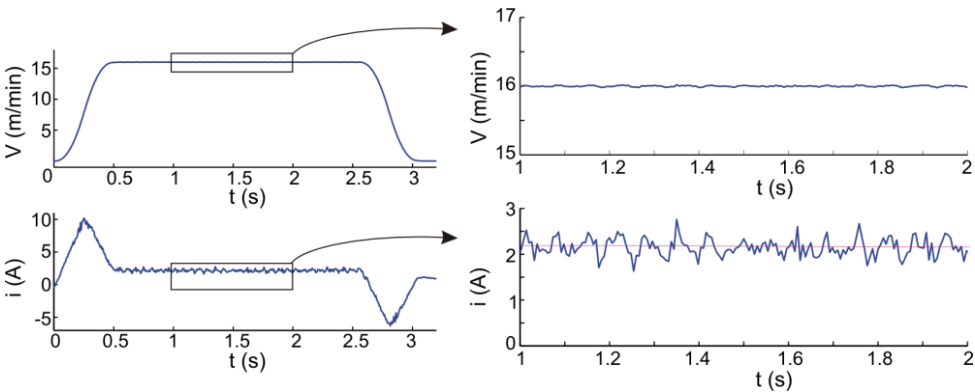

Hence, the law $i_{friction} = f(V)$ is established by fitting a trend curve to experimental points (**Figure 5**). Measurements are performed for both positive and negative displacements along the axis. As results are quite similar, a symmetrical model around the origin is chosen to fit the current.

**Figure 5** Friction law and experimental results (X axis)

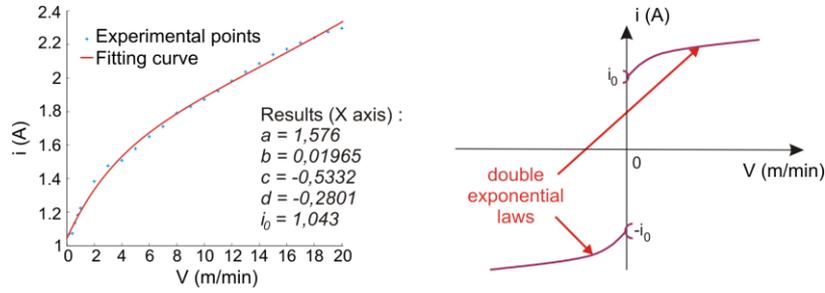

The chosen model is a double exponential law that gives a regression coefficient of 0,995. The final model is given by equations:

$$\begin{cases} i_{friction}(V) = a \cdot e^{b.V} + c \cdot e^{d.V} & if\ V \in ]0;20] \\ i_{friction}(V) = -a \cdot e^{-b.V} - c \cdot e^{-d.V} & if\ V \in [-20;0[ \\ i_{friction} \in [-i_0;i_0] & if\ V = 0 \end{cases} \quad (3)$$

## 3.4 Equivalent inertia and masses

Concerning the motor parameters, electric parameters can be found in manufacturer's documentation. An experimental identification is performed for mechanical parameters, which account for embedded masses and other inertias. Equivalent inertia relative to the motor shaft is calculated from a series of current measurements with zero acceleration displacements according to equation (5). $J_{eq}$ is then calculated for each time step and an average is computed on each time interval for which acceleration is non zero.

$$J_{eq} = \frac{K_t \cdot i - C_r}{\frac{d\Omega_m}{dt}} \quad (5)$$

As previously suggested, gravity is taken into account, especially for vertical axes and tilting axes, by adding resistant torques. These ones are easily determined with a measurement of the current on static positions, when there is neither velocity nor acceleration.

## 3.5 Feedforward and shift parameters

Feedforward is a common technique used to cancel dragging differences on the position (Lambrechts et al., 2005). Two types of feedforward are commonly implemented and will be modelled bellow: the torque feedforward and the velocity feedforward, which can be disabled or used separately or together. Therefore, constants associated to the feedforward compensations can be evaluated from the measurement of the feedforward setpoints TFFWS and VFFWS.

The ratios between the first and second derivatives of the Position Setpoint at Entrance of the Controller (PSEC) and VFFWS or TFFWS respectively are supposed to be constant:

$$VFFWS(t) = VFFW \cdot \frac{d(PSEC(t))}{dt} \quad (6)$$

$$TFFWS(t) = TFFW \cdot \frac{d^2(PSEC(t))}{dt^2} \quad (7)$$

Derivatives are computed according to the Euler approximation (with $z^{-1}$ the backward shift operator and $Tps$ the sampling rate):

$$\begin{aligned} y(t) = \frac{d(x(t))}{dt} &\leftrightarrow Y(z) = \frac{1 - z^{-1}}{Tps} \cdot X(z) \\ u(t) = \frac{d^2(x(t))}{dt^2} &\leftrightarrow U(z) = \frac{1 - 2 \cdot z^{-1} + z^{-2}}{Tps^2} \cdot X(z) \end{aligned} \quad (8)$$

Then, a linear regression is applied to measurements to verify that feedforward terms are constant.

Following the identification stage, an adjustment of the parameters is performed. To do this, three parameters (α, β, γ) are introduced (see Appendix 1), which stand for a delay in the feed drive structure. They are optimised using a least-square minimisation of the deviations between simulated and measured positions. To begin with, the first parameter α located in the position loop is identified without feedforward actions. Next, β is identified in the same way with only the velocity feedforward and finally, the last one γ when all anticipations are activated. Values of adjustment parameters are given in Appendix 2.

# 4 Model assessment

This section details the model assessment for the 5-axis milling centre Mikron UCP710 performed thanks to trials. Different types of trajectories are tested, involving one or several axes. Data are directly recorded via the NC oscilloscope at the position loop sampling period. The measured signals are the position setpoints (PSEC), the actual position recorded through the encoders (mm), the current (A) and the velocity (m/min). Collected data are directly confronted to values issued from simulations.

## 4.1 One axis validation

Testing one-axis permits to check independently each axis while being free from coupling problems. Trials are performed for translational axes as well as rotational ones. Whatever the axis, all the experiments lead to similar results. Therefore, only tests concerning the X translational axis and the C rotational axis are presented in the paper. Considering a segment defined from two position values, several tests are conducted:

- Case 1: the programmed velocity is constant during the displacement (**Figure 6**)
- Case 2: the programmed velocity is V1 for a segment portion, and is V2 for the other part, with V1>V2 (**Figure 7**)
- Case 3: the segment is travelled in back and forth at the same programmed velocity.

The case 1 is illustrated for the X axis in Figure 6. Simulated values are closed to the measured ones as deviations do not exceed 4 µm except during critical phases of acceleration and deceleration. In such cases, differences can reach 6 µm. Simulated and measured values for the velocity and current are almost identical. Hence, values are small enough to assess the model of each translational axis independently, with or without feedforward actions.

Case 2 is illustrated for the rotational C axis in Figure 7. Practically, this test consists in covering the angular interval [0° , 130°] at the speed of 18 rpm and the interval [130° , 210°] at 6 rpm. Discrepancies collected between angular positions measured and simulated are smaller than $6.10^{-3}$ degrees for this axis, which is rather a good result. Once more, the maximum deviations are obtained during deceleration and acceleration phases.

**Figure 6** X axis translation test

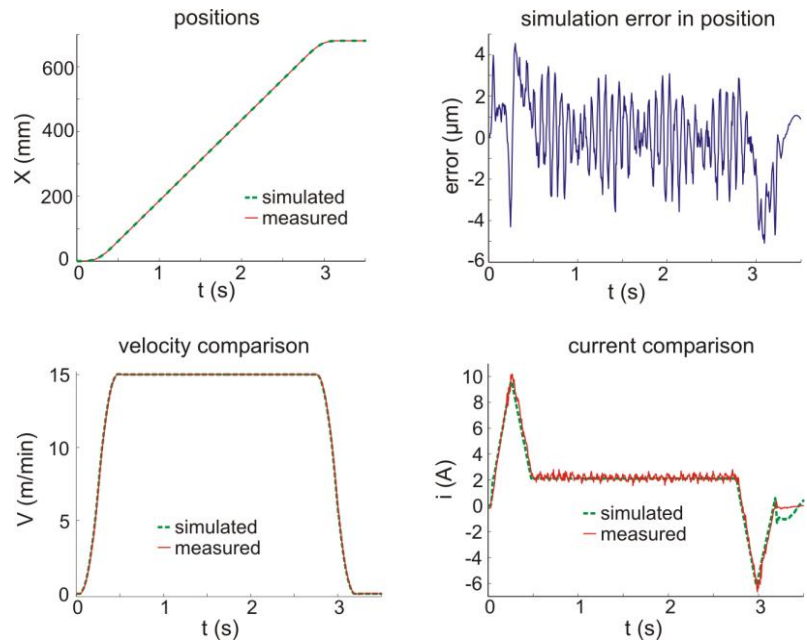

**Figure 7** Two velocities rotation test

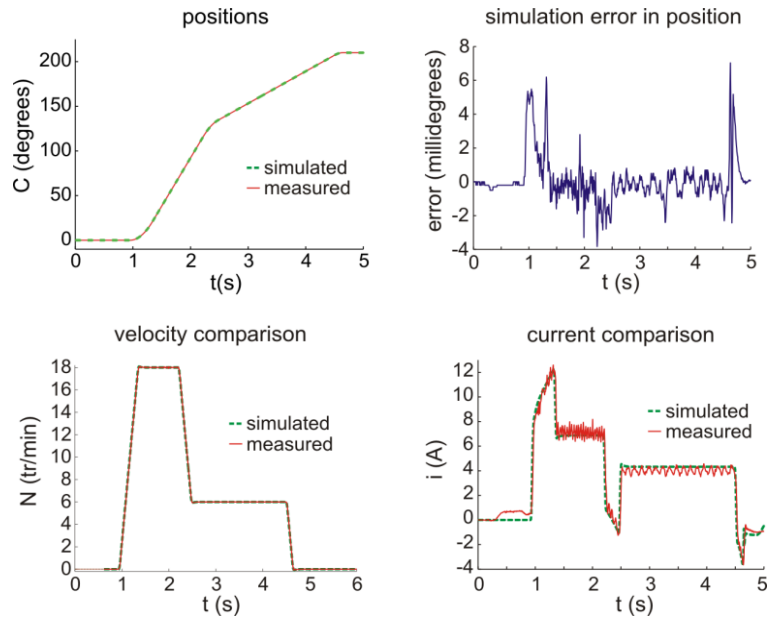

Hence, whatever the axis nature (translational or rotational), and the motorization (synchronous rotational or synchronous linear motors), differences between measurements and simulations are sufficiently small to assess the model for each axis independently solicited. The accuracy of the simulation (globally less than 0.005 mm or 0.005°) seems to be enough to predict tracking deviations.

*4.2 Multi axis tests*

These tests aim at the model assessment for trajectories involving several axes at the same time. Moreover, trajectories present some geometrical singularities implying significant tracking deviations in particular for high velocities.

The four trajectories tested are presented in Figure 8. Figure 8.a and 8.b report results concerning trajectories executed in the X-Y plane, with a programmed feedrate Vf =10 m/min. Figure 8.c stands for tool orientation change in the part coordinate system considering a trajectory defined in the A-C plane. Finally, in Figure 8.d, the trajectory is a 3D B-Spline curve.

For each case, differences between simulated and measured values are estimated (right part of the figures). Deviations between simulations and measurements never exceed 0.01 mm for translational axes, respectively 0.01° for rotational axes; deviations are below 5 µm (resp. 0.005°) for all the trials. However, the deviations reach 12 µm for the 3D curve. Such differences are due to chatters that are discernible during acceleration or deceleration phases when the trajectory is executed on the machine tool.

Differences observed between simulated and measured data are likely due to the difficulty of modelling the specific numerical and software NC treatments. Indeed, such elements are considered through the adjustment parameters (a, b, g) which are not so simple to precisely identify. Furthermore, accounting for axis coupling or backlash compensations could also improve the model precision.

Nevertheless, the comparisons proposed in Figure 8 enhance the good behaviour of the model to simulate trajectory tracking whatever the trajectory executed. Moreover, the model allows the prediction of tracking deviations that can impact the geometrical quality of the parts (Prévost et al., 2008).

**Figure 8** Multi axis tests

In the following, one application using the model presented above is studied. It deals with the implementation and the tuning of advanced controllers, and more particularly predictive controllers based on a GPC structure.

## 5 Towards an optimized RST Generalized Predictive Control

This section presents the implementation of an advanced controller structure based on the GPC algorithm. Among various strategies of advanced controllers, previous research works in the field of machine tool in particular show that GPC can provide better tracking performance compared with classical cascaded structures (***Susanu et al., 2004).

However, these works also highlight that the choice of the required tuning parameters is a crucial point to obtain good tool path tracking performance, which may depends not only on the behavior of the controlled loops, but also on the kind of trajectory issued from the interpolation module. On the other hand, implementing predictive strategies is performed with the help of a prediction model of the system, which is usually selected as simple as possible. These reasons are strong motivations in favor of the model developed in this paper, which will be very helpful in particular from a tuning point of view. This aspect will be further examined in the specific case of the Mikron UCP 710 machining centre.

A first paragraph summarizes the basic ideas of predictive control, leading to the elaboration of the predictive controller under the RST structure. A second paragraph then details the implementation of such a strategy in the case of trajectories generated through circular interpolation, emphasizing the need for optimization of the parameters depending on the type of trajectory.

### 5.1 GPC strategy and implementation of the new structure

Based on Model Predictive Control philosophy, the Generalized Predictive Control strategy relies on four important ideas reproducing the basic decisional of the human behaviour (Rossiter, 2003), creation of an anticipative effect by exploiting the trajectory to be followed in the future, definition of a numerical representation of the process as the prediction model, minimization over a finite horizon of an objective function elaborating the best control action, time-domain displacement of predictions known as receding horizon strategy. These basic steps can be translated in equations and algorithms which are summarized in (Susanu et al., 2006), leading to a controller finally given in the so-called RST formalism. These polynomials R, S and T are completely determined by algorithms that will not be detailed here and whose coefficients derived from the resolution of Diophantine equations. As a result, the tuning of four parameters $N_1$, $N_2$, $N_u$ and $\lambda$ (see Table 1) is required prior to the elaboration of the three polynomials R, S and T, which are then computed offline and uniquely defined.

**Table 1** RST tuning parameters

| | |
|---|---|
| $N_1$ | Minimum output prediction horizon |
| $N_2$ | Maximum output prediction horizon |
| $N_u$ | Control horizon |
| $\lambda$ | Weighting factor |

With this structure, the T polynomial has a non causal form, thus ensuring an anticipative effect in closed-loop. Consequently, there is no need anymore for implementing feedforward actions which are difficult to tune, as usually done in classical NC. Finally the RST controller is implemented as shown in Figure 9. In the following test, the RST formalism is implemented in a cascaded version, only for the position loop of each axis drive.

**Figure 9** The RST polynomial structure

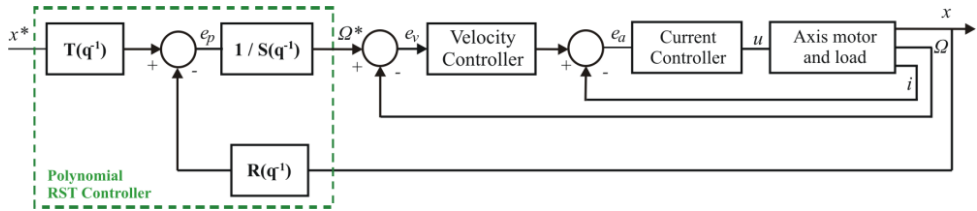

Feedback is thus carried out under a numerical RST structure, only requiring the computation of a simple difference equation:

$$S(q^{-1})\Omega^*(t) = -R(q^{-1})x(t) + T(q^{-1})x^*(t) \qquad (9)$$

where $x^*$ is the input, $x$ the output and $\Omega^*$ the velocity reference signal.

Remark 1: Using this RST formalism to structure the axis control law represents an important advantage since this form appears to be very generic. Indeed, a large number of numerical controllers can be structured in the same way, from basic PID to more advanced ones. Within the framework of open architecture, axis controllers can thus be seen as plug and play modules, ensuring interchangeability features.

Remark 2: Another interest of this generic structure is the resulting reduced computation time as the controller polynomials are in most cases of small degrees. This aspect allows the implementation of this strategy even with short sampling times, as for instance in the case of very high speed machining.

Based on the feed drive model developed in the previous sections, the following section presents the results of simulation with the predictive RST control structure on the position loop, considering a tuning of the four parameters $N_1$, $N_2$, $N_u$ and $\lambda$ ensuring good stability and robustness properties (***Susanu et al., 2004). Comparisons are made between classical control architecture with feedforward terms and predictive RST control structure for two types of trajectories.

## 5.2 RST simulations and comparisons with the classical control architecture

The first case consists in plane circular trajectories, so-called ballbar test, largely used to analyse and measure tracking errors and geometrical errors of the machine tool. A circular trajectory of radius 150mm is first considered, with a programmed feedrate of 15m/min. The trajectory is interpolated using circular interpolation (G2/G3) and with velocity and torque feedforward actions on. Figure 10 represents tracking errors with classical cascaded control structure and feedforward actions (left) and predictive RST control structure (right).

**Figure 10** Tracking error comparison between cascaded and RST structure – circular trajectory

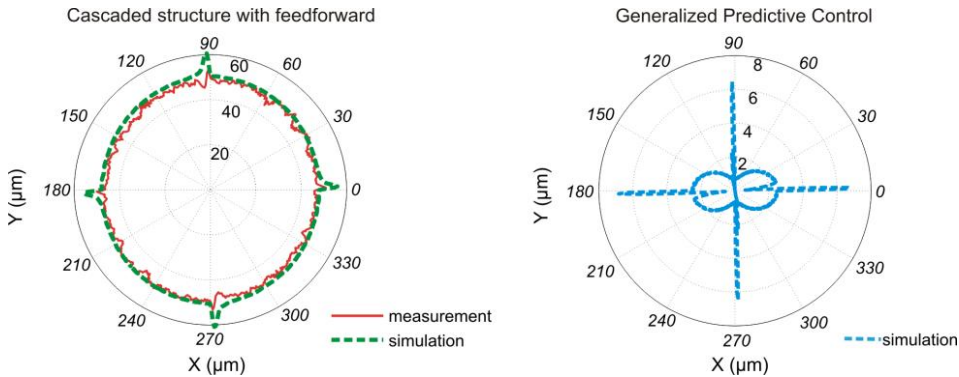

It can be clearly seen that the 50mm tracking error observed with the cascaded structure becomes almost zero with the RST structure. As a first conclusion, the chosen parameters tuning of the GPC structure greatly improves the axes behaviours. The fact that the industrial NC module is completely closed does not enable to validate the predictive law through real experiments. In that sense, the elaboration of a realistic and simple model as proposed in the paper is a major advantage which helps validating new advanced controllers.

The second case, for which trajectories present high tangency discontinuities, is now studied. For example, two trajectories in the X-Y plane represented on Figure 11 are tested. The geometries chosen here are sharp corners (45° and 90°), which generally induce high deviations on machined parts with high programmed feedrates. The programmed feedrate speed for each case is 10m/min. Figure 11 presents the behaviour of the simulated trajectories, obtained for the predictive controller with the same tuning as for the first case.

**Figure 11** Trajectory comparison between cascaded and RST structure – sharp corners

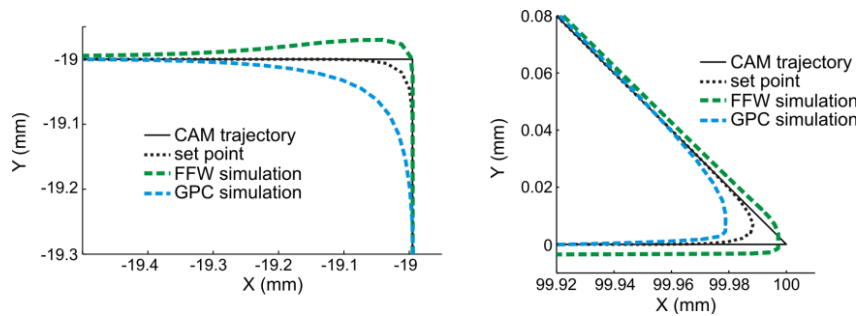

A decrease of the gap between simulated position with the new control strategy (GPC) and the position setpoint is illustrated on the right handside of the figure. Unfortunately, this is not the case for the left handside of this figure. This case highlights the need for optimization of the tuning of the four parameters $N_1$, $N_2$, $N_u$ and $\lambda$. This can be realized using the model developed in this paper. Indeed, the virtual environment developed around the feed drive model is a useful tool to select through simulation the adequate tuning for the set of parameters, in particular in function of the kind of trajectory that has to be followed. One perspective is thus to find an optimisation technique and propose such a technique in function of the tool path typology.

## 6 Conclusion

This paper focuses on a model of the feed drives of machine tool. The proposed model is deliberately simple to be applied to milling centers equipped with modern industrial CNC to simulate tracking errors. The model accounts for the main influent components on the dynamical axis behaviour: inertia, friction, and feedforward compensations. Complex phenomena as for specific NC treatments are taken into account through adjustment parameters. An original procedure of parameter identification is carried out thanks to the capture on fly of various variables using the oscilloscope of the NC unit. Therefore, the virtual environment developed around the feed drive model gives a useful tool to simulate tracking errors, but also to investigate various control structures.

The model is implemented with success for a 5-axis milling center. Deviations observed between the measured displacements and the simulated ones relatively low assess the relevancy of the model. To enhance the benefit of such a model various types of trajectories are tested. Results clearly emphasize the good behaviour of the model to simulate trajectory tracking whatever the trajectory executed, even for 3D trajectories. The model turns out to be useful for the prediction of tracking deviations that can impact the geometrical quality of the parts.

A second major interest of the model is its adaptability to various control structures. In this direction a Generalized Predictive Control is implemented into the model to design a new framework of control. From the control point of view, the elaboration of a virtual environment based on a simple but realistic model of the feed drives presents at least two important advantages. First, it allows testing and validating through realistic simulations new advanced control architectures, which would not have been possible up to now on industrial machining centers due to the closed architecture of the NC modules. Then, it allows through several simulations selecting the best set of tuning parameters to provide stability and robustness in relation with a specific trajectory.

## Acknowledgements

Authors would like to thank AIP Primeca Ile de France who have graciously allowed to use of the 3x DMG HSC75 milling center.
This work was carried out within the Manufacturing 21 working group, which comprises 18 French laboratories. The topics approached are:
- modelling,
- virtual machining,
- emergence of new manufacturing methods.

**Appendix 1**

# Appendix 2

| Name | Description | Units | 5x Mikron UCP 710 values | | |
|---|---|---|---|---|---|
| | | | X axis | Y axis | Z axis |
| $J_{eq}$ | Equivalent inertia | kg.m² | 0.028 | 0.024 | 0.0225 |
| $K_P$ | Position controller gain | m/(min/mm) | 1.5 | 1.5 | 3.5 |
| $K_V$ | Speed controller gain | N.m/(rad/s) | 5.0 | 4.5 | 5.5 |
| $T_V$ | Speed controller integration time | ms | 4 | 6 | 4 |
| $K_I$ | Current controller gain | V/A | 13.0 | 12.0 | 12.0 |
| $T_I$ | Current controller integration time | ms | 2 | 2 | 2 |
| Tsp | Position controller cycle time | ms | 6.0 | 6.0 | 6.0 ms |
| Tsv | Velocity controller cycle time | μs | 250 | 250 | 250 |
| Tsi | Current controller cycle time | μs | 125 | 125 | 125 |
| A | First friction model parameter | A.min/m | 1.576 | 1.253 | 1.420 |
| B | Second friction model parameter | | 0.01965 | 0.01895 | 0.01650 |
| C | Third friction model parameter | A.min/m | -0.5332 | -0.3629 | -0.6301 |
| D | Fourth friction model parameter | | -0.2801 | -0.4026 | -0.2625 |
| $i_0$ | Coulomb friction intensity equivalent | A | 1.043 | 0.890 | 0.790 |
| α | Adjustment parameter on position loop | ms | 9 | 9 | 9 |
| β | Adjustment parameter on velocity feedforward | ms | 9 | 9 | 9 |
| γ | Adjustment parameter on torque feedforward | ms | 9 | 9 | 9 |
| VFFW | Velocity FeedForward constant | | 1 | 1 | 1 |
| TFFW | Torque FeedForward constant | kg.m² | 0.002034 | 0.002030 | 0.002030 |